\documentclass[aps,prb,preprintnumbers,showpacs]{revtex4}
\usepackage{graphicx}
\usepackage{bm}
\begin{document}
\newcommand{\beq}{\begin{equation}}
\newcommand{\eeq}{\end{equation}}
\newcommand{\bea}{\begin{eqnarray}}
\newcommand{\eea}{\end{eqnarray}}
\def\plumin{\underline{+}}
\def\minplu{{{\stackrel{\underline{\ \ }}{+}}}}
\def\br{{\bf r}}
\def\bR{{\bf R}}
\def\bfr{{\bf r}}
\def\bfk{{\bf k}}
\def\bfq{{\bf q}}
\def\bq{{\bf q}}
\def\brho{\bf{\rho}}

\title{
Nonlinear screening and stopping power in two-dimensional 
electron gases }
\author{E. Zaremba}
\affiliation{
Department of Physics,
Queen's University,
Kingston, Ontario, Canada K7L 3N6 and
Donostia International Physics Center (DIPC), Paseo Manuel de 
Lardizabal, no. 4, 20018 San Sebatian, Basque Country, Spain
}
\author{I. Nagy}
\affiliation{
Department of Theoretical Physics, Institute of Physics, 
Technical University of Budapest, H-1521 Budapest, Hungary, and
Donostia International Physics Center (DIPC), Paseo Manuel de Lardizabal, no.
4, 20018 San Sebatian, Basque Country, Spain}
\author{P. M. Echenique}
\affiliation{
Departamento de Fisica de Materiales,
Facultad de Quimica, Universidad del Pais Vasco
and Centro Mixto CSIC-UPV/EHU, 
Apdo. 1072, 20018 San Sebastian, Basque Country, Spain 
}

\date{\today}


\begin{abstract}
We have used density functional theory to study the nonlinear screening
properties of a two-dimensional (2D) electron 
gas. In particular, we consider the screening of an external static
point charge of magnitude $Z$ as a function of the distance of the
charge from the plane of the gas. The self-consistent screening
potentials are then used to determine the 2D stopping power in the low
velocity limit based on the momentum transfer cross-section.
Calculations as a function of $Z$ establish the limits of validity of
linear and quadratic response theory calculations, and show that
nonlinear screening theory already provides significant corrections in 
the case of protons. In contrast to the 3D situation, we find that
the nonlinearly screened potential supports a bound state even in the
high density limit. This behaviour is elucidated with the derivation of
a high density screening theorem which proves that the screening charge
can be calculated perturbatively in the high density limit for arbitrary
dimensions. However, the theorem has particularly interesting
implications in 2D where, contrary to expectations, we find that
perturbation theory remains valid
even when the perturbing potential supports bound states. 
\end{abstract}

\pacs{PACS Numbers: 73.20.-r, 73.20.Hb}
\maketitle

\section{Introduction}

The screening response is a 
fundamental property of an electron gas in arbitrary dimensions. 
The situation in two dimensions is of particular interest
because of the possible realization of quasi-two-dimensional
systems in semiconductor heterostructures\cite{ando82}, image and 
band-gap surface
states at metal surfaces\cite{echenique04}, electrons at the surface of
liquid helium\cite{andrei97}, metallic overlayers
on insulating substrates and layered materials\cite{layered}. 
Examples of problems in which an understanding of the screening of
an external charge is important include impurity-limited electron 
transport in two-dimensional electron gases\cite{ando82}, the electronic 
structure of intrinsic or photo-induced defects in semiconductor
heterostructures\cite{fletcher90}
and dynamic interactions with moving charges\cite{nagao01,murphy95}. 
In each of these
cases, the strong interaction of mobile electrons with the charged 
impurity leads to significant modifications of the local electronic 
structure.

One of the first attempts to deal with 2D screening was the work of
Stern and Howard\cite{stern67} which was motivated by measurements of
electron mobilities in Si inversion layers. To perform explicit
calculations they considered a model in which the two-dimensional
electron gas (2DEG) is represented as a sheet of zero thickness,
the ideal 2D limit, with a
charged impurity situated a distance $d$ from the sheet. The screened
impurity potential was then obtained within a Thomas-Fermi
approximation and was used to investigate the possibility of impurity
bound states. However no attempt was made to perform fully
self-consistent calculations of the screening charge density and the
screened potential. This was taken up later in a series of papers by 
Vinter\cite{vinter82} for a Si inversion layer. 
This work is especially notable for its attempt to realistically
account for the underlying bandstructure of the semiconductor and the 
finite thickness of the 2DEG. A detailed discussion of screening and
bound state formation was presented for this particular heterostructure
and the results were shown to be consistent with observed 
impurity-limited mobilities. 

Also of interest is the dependence of screening on the dimensionality 
of the system. That important differences
might arise is indicated by the well known fact that any
purely attractive potential in 2D always has at least one bound
state\cite{simon76}, in sharp contrast with the situation in 3D.
Extending previous nonlinear screening calculations
from 3D\cite{zaremba77} to 2D and
exploring these possible differences is part of the motivation 
for the work described in this paper. To do this we perform
self-consistent nonlinear screening calculations for an ideal 2DEG
within the context of density functional theory. Our studies as a
function of density reveal some interesting properties of the screening
in 2D that had not been appreciated before. The interpretation of  
these numerical results is facilitated by the proof of a general 
theorem regarding the nature of screening in the high density limit.
A preliminary account of some of this work was presented
earlier\cite{zaremba03}.

As a specific application of the results we consider the low velocity stopping
power for a heavy projectile moving parallel to the plane of the 2DEG.
This problem has been treated previously in both linear and nonlinear
response approximations. Of the linear theories, we mention
the early work of Fetter\cite{fetter73} using a hydrodynamic model which 
was followed by a calculation based on the random phase
approximation (RPA) by Horing {\it et al.}\cite{horing87}.
In this latter work a finite distance, $d$, between the projectile and the 
plane of the gas was considered. The case of a
projectile moving in the plane of the gas has also been studied, not only
within the RPA\cite{bergara97}, but also with the inclusion of local field
corrections\cite{wang95} and for a gas at finite temperatures\cite{bret93}.
Of course the interaction cannot always be considered as weak and a 
fully nonlinear theory of the screening is in general necessary. 
One step in this direction is the RPA quadratic response 
formulation\cite{hu88}
which was used by Bergara {\it et al.}\cite{bergara99} in 2D to 
provide the lowest order nonlinear correction to the stopping 
power. Although this theory is perturbative and therefore
limited in its range of validity, it 
has the merit that the nonlinear correction is determined
for arbitrary projectile velocities. 

The stopping power at low projectile velocities can also be 
formulated in terms of the scattering of electrons from the
screened potential. 
For qualitative purposes it is sometimes useful to consider model
potentials\cite{nagy95} whose arbitrariness can be limited by
imposing physical constraints such as the Friedel sum rule. 
This was the approach taken by Wang and
Ma\cite{wang97} to determine proton and antiproton stopping powers in a
2DEG. A somewhat more fundamental approach based on the nonlinear
screening theory of Sj\"olander and Stott\cite{sjolander72} was used by
Krakovsky and Percus\cite{krakovsky95} for negatively charged 
projectiles. Their results for this case are consistent with the model 
potential results of Wang and Ma\cite{wang97}. None of these
calculations, however, are truly self-consistent. Our results based on
self-consistent density functional theory calculations eliminate
this source of uncertainty in the determination of 2D stopping powers. 

\section{Self-consistent Screening}

The electrons of our ideal 2DEG are assumed to have 
an isotropic effective mass $m^*$ and move in the presence of a uniform
neutralizing positive background. We also imagine the entire 2D system
to be immersed in an extended dielectric with permittivity 
$\varepsilon$ as might occur in a semiconductor heterostructure.
Other dielectric arrangements can also be envisaged, for example the
surface of liquid helium, but these will not be considered here.
All lengths will be expressed in units of the effective Bohr radius
$a_0 = \varepsilon \hbar^2/m^*e^2$, and all energies in units of the
effective Hartree, $H = e^2/\varepsilon a_0$. The mean density
of the gas, $n_0$, is characterized by the density parameter
$r_s = 1/\sqrt{\pi n_0}$.

Our objective is to determine the nonlinear screening of a
stationary point charge, $Z$, located a distance $d$ from the plane of
the 2D gas. The methodology follows that used in earlier calculations 
for a three-dimensional gas\cite{zaremba77,nagy89}
and is suitable for both positive and negative charges.
This screening response is determined by solving
self-consistently the two-dimensional Kohn-Sham equations
\beq
-{1\over 2} \nabla^2 \psi_i(\bfr) + \Delta v_{\rm eff}(\bfr) 
\psi_i(\bfr)= E_i \psi_i(\bfr) 
\label{SE}
\eeq
where the effective potential is given by
\beq
\Delta v_{\rm eff}(\bfr) = v_{ext}(\bfr) + \Delta v_H(\bfr)+
\Delta v_{xc}(\bfr)\,.
\label{veff}
\eeq
The external potential $v_{ext}(\bfr)$ in the plane of the gas is
$-Z/\sqrt{d^2 + r^2}$, and $\Delta v_H(\bfr)$ is the Hartree potential 
\beq
\Delta v_H(\bfr) = \int d^2r' {\Delta n(\bfr') \over |\bfr-\bfr'|}
\label{Hartree}
\eeq
due to the electronic screening density, $\Delta n(\bfr) = n(\bfr)-n_0$.
The change in the exchange-correlation potential, $\Delta v_{xc}(\bfr) =
v_{xc}[n(\bfr)] - v_{xc}[n_0]$, is 
defined in the local density approximation (LDA)
using the parameterization of the 2D exchange-correlation energy
given by Tanatar and Ceperley\cite{tanatar89}. 

The total screening density is given by
\beq
\Delta n(\bfr) = \sum_b |\psi_b(\bfr)|^2 + \sum_i \left 
[|\psi_i(\bfr)|^2 - |\psi_i^0(\bfr)|^2 \right ]
\eeq
where the first sum extends over all bound states of the effective
potential, and the second extends over all occupied continuum states up
to the Fermi level $E_F$. We assume that each spatial orbital is
doubly-occupied for spin. The free-particle solutions
$\psi_i^0(\bfr)$ are obtained in
the absense of the external potential. Due to the cylindrical
symmetry of the problem, the solutions of Eq.~(\ref{SE}) have the
form $\psi_i(\bfr) \equiv e^{im\phi} R_{nm}(r)$ where the
angular momentum quantum number $m$ takes on integral values and
$n$ distinguishes the different radial
solutions\cite{stern67,adhikari86}. The continuum states 
behave asymptotically as $R_{km}(r) \sim r^{-1/2}
\cos(kr -{1\over 2}|m|\pi - {1\over 4}\pi+ \eta_m)$ where $k =
\sqrt{2E}$ is the wavevector and $\eta_m(k)$ is the 2D scattering phase 
shift. For the free-particle solutions, $\eta_m = 0$. These
scattering phase shifts are related to the total screening density 
according to the 2D Friedel sum rule (FSR)\cite{stern67}
\beq
 Z_{FSR} = {2\over \pi}\sum_{m=-\infty}^\infty \eta_m(E_F)\,.
\label{FSR}
\eeq
At self-consistency, this quantity must equal the charge $Z$ of the 
external impurity.

The solutions of Eq.~(\ref{SE}) are obtained using standard
numerical techniques. The calculation of $\Delta v_H(\bfr)$, the only
nontrivial step in determining the effective potential, is 
performed by
making use of a fast fourier transform technique to go successively
between real and wavevector space. Details of the method
are given in an Appendix.

\section{Numerical RESULTS}

Self-consistent calculations were performed for a range of 
densities and impurity charges $Z$, including negative values. In
addition, calculations were done with and without exchange-correlation
effects included in order to compare with earlier linear calculations
which were performed within the random phase approximation (RPA).

\subsection{$Z = 1$, $d=0$}

To begin we show results for a charge $Z= 1$ in the plane of the 
2DEG ($d=0$). Fig.~1 presents the self-consistent effective potential 
$\Delta v_{\rm eff}$ for $r_s$ values ranging between 1 and
10 in steps of 1. This range spans low to high densities and in
particular includes densities of experimental interest. 
Also shown for comparison is the Thomas-Fermi (TF) screened
potential which, as will be explained shortly, is the $r_s \to 0$
limit of the DFT calculations. As such, we see that the potential for
$r_s= 1$ is already quite close to the TF limit. 

The variation with
$r_s$ is systematic, evolving with increasing $r_s$ into a potential 
with a strongly repulsive
region beyond $r \simeq 0.5$ a.u. This behaviour can be explained in
terms of the
underlying electronic structure of the screened impurity. For all the 
densities shown, the screened potential supports one bound state which
is occupied by two electrons. Since the total screening density must
integrate to unity according to the FSR in Eq.~(\ref{FSR}), 
the continuum states must themselves contribute a total
charge of $+1$
in order to compensate for the overscreening provided by the two
bound electrons. This is illustrated in Fig.~2 which shows the bound and
continuum densities for two representative $r_s$ values, $r_s =2$ and
10. It can be seen that the bound state distribution is very similar 
despite the very different background densities.
This is consistent with the similarity of
the inner region of the potentials in Fig.~1 which is primarily
responsible for the shape of the bound state. At large distances from
the impurity the densities in Fig.~2 exhibit characteristic Friedel
oscillations with wavelength $\pi/k_F$.

The negative portion of the screening densities in Fig.~2 corresponds to a
local charge density which is positive and provides the necessary 
compensation of
the negative bound state electronic charge. However, the distribution in 
space of this compensating charge is quite different in the two cases and 
leads to the different behaviour of the $r_s =2$ and $r_s
=10$ potentials for $r > 0.5$ a.u. The form of the $r_s = 10$ continuum
density is particularly interesting. In the range $ 0 \le r \simeq 10$
a.u., $\Delta n_{\rm cont}(r)$ has approximately a constant value of
$-n_0$ (hence $r\Delta n_{\rm cont}(r)$ behaves linearly
as seen in Fig.~2). In other words, the $Z=1$ impurity with 2 bound
electrons can be viewed as an H$^-$ ion which sits at the center of a
positively charged disc of radius $R \simeq r_s$.
Returning to Fig.~1, we can now understand the repulsive part of the
potentials seen for large $r_s$. The LDA potential of a 2D H$^-$ ion 
itself has a $+1/r$ Coulomb tail arising from the Hartree
potential. In the 2DEG environment, this H$^-$ potential
is screened by the positive background and as a result, the potential
goes to zero at approximately $R \simeq r_s$.

Another interesting quantity is the bound state
energy shown in Fig.~3 as a function of $r_s$. The
solid curve is the result of the full DFT calculation including xc,
while the dashed curve is the corresponding result obtained when 
$\Delta v_{xc}$ is set to zero.
We refer to the latter as the Hartree approximation. The
Hartree result is monotonically decreasing with decreasing $r_s$ and
goes to a limiting value
of $E_0 = - 0.2862$ $H$ at $r_s = 0$ which, as we shall see, is the
binding energy for the TF screened potential. The xc result likewise
decreases with decreasing $r_s$, but reaches a mininum near 
$r_s \simeq 0.5$ and then increases to the same limiting value at 
$r_s =0$ as in the Hartree approximation. The more negative xc 
eigenvalue reflects the stronger binding
due to the xc potential which is attractive in the vicinity of the
positive impurity as a result of the pile-up of the electron
screening density.
To illustrate this we compare the self-consistent potentials
with and without xc for $r_s=2$ in Fig.~4. It can be seen from this
figure that $\Delta v_{\rm eff}$ with xc is more attractive in the core 
region which
is important for the determination of the bound state. Associated with
this is a relative phase shift of the long range Friedel
oscillations between the two cases.

The decreasing trend in the xc eigenvalue is consistent with the
behaviour of the potentials in Fig.~1 but is opposite to what
is found in the analogous 3D calculations\cite{zaremba77}.
There the bound state energy {\it increases} with decreasing $r_s$, 
that is, the bound state becomes shallower with increasing
density.  The explanation for the 3D behaviour is that the screening of
the impurity potential is more effective with increasing density,
and as a result, the bound state eventually ceases to exist
beyond a certain critical density. This behaviour is consistent with the
expected
applicability of perturbation theory in the high density limit when the
Fermi energy $E_F$ is much larger than the magnitude of the screened 
potential $|\Delta v_{\rm eff}(r)|$\cite{footnote2}. If one were to use
the same reasoning in 2D, the screening density in
wavevector space would be given by the linear response result 
(Sec. IV provides further details)
\beq
\Delta n(q) = \chi_0(q) \Delta v_{\rm eff}(q) \,,
\eeq
where $\chi_0(q)$ is the Fourier transform of the static noninteracting
density response function of the 2D gas. Within linear response theory,
the screening density arises from the perturbation of the plane wave 
states of 
the uniform gas and there cannot be a bound state contribution. 
However, as we have just seen, bound states do persist in 2D even in the 
high density limit. This is a peculiarity of 2D and is associated with the 
fact that a purely {\it attractive}
potential always has at least one bound state 
regardless of the strength of the interaction\cite{simon76}.
This effect is obviously missed by linear
response theory and would seem to invalidate a perturbative approach.
Nevertheless, it turns out that the results of linear 
response theory are indeed correct
even though the theory neglects bound states entirely. We
defer discussion of this point to the following Section.

   The opposite limit $r_s \to \infty$ is also of interest. In the
Hartree approximation, the bound state eigenvalue approaches zero near $r_s
= 8$. Whether a very shallow bound state persists beyond this value 
could not be confirmed due to difficulties in
obtaining converged self-consistent solutions. This was less of a
problem with the inclusion of exchange and correlation 
since these additional interactions stabilize the
bound state relative to the Hartree calculation. In the range $8 \le r_s
\le 12$ the bound state eigenvalue could be fit approximately to the 
expression $E_0 
\simeq 0.023 - 0.7/r_s$ $H$. Extrapolation of the numerical data beyond
$r_s = 12$ suggests that the bound state eigenvalue goes to zero at some 
large, but finite, $r_s$ value. However, it was not possible to confirm
this since it was difficult to obtain converged solutions for large
values of $r_s$. Thus we
cannot state unequivocally what the large $r_s$ behaviour is, and in
particular, whether or not a free H$^-$ ion is stable in 2D within the
LDA. In 3D it is known\cite{shore77} 
that the H$^-$ ion is {\it not} stable within the
LDA and our data suggest that this might also be the case in 2D. This
remains an interesting question to explore in the future.

As a final comment about the eigenvalue obtained with xc, we note that
the H$^-$ ion is stabilized in the range studied by the fact that it
sits at the center of a positively charged disc of radius $R \simeq
r_s$. The electrostatic potential due to this charge has
the value $-2/r_s$ $H$ at the centre of the disc relative to infinity. 
The depletion of charge
also gives rise to a shift in the xc potential of $-v_{xc}(n_0)$ which
is {\it positive} and therefore a destabilizing effect. For large $r_s$,
$-v_{xc}(n_0) \simeq +1.6/r_s$ $H$ which, when
combined with the electrostatic shift, gives a potential shift of about
$-0.4/r_s$ $H$. It is clear that this represents a significant contribution
to the xc eigenvalue at the larger $r_s$ values we have considered. However 
the existance of a bound state cannot be determined without a detailed 
knowledge of the
self-consistent potential near the impurity. In the case of the Hartree
approximation, the 
stabilizing effect of the electrostatic potential of the positively 
charged disc ($-2/r_s$ $H$) is clearly insufficient to compensate
for the destabilizing effect of the Hartree interaction between the two
bound electrons in the large $r_s$ limit.

\subsection{$Z= -1$, $d=0$}

We consider next the situation of a negatively charged impurity such as
an antiproton or acceptor state in a semiconductor\cite{richter89}. 
Fig.~5 illustrates
the variation of the screening density with $r_s$. The results
found here in 2D are similar to those found previously in
3D\cite{nagy89}. The
negative impurity repels electrons from its vicinity and roughly
speaking exposes a positively charged disc of radius $R \simeq r_s$
which neutralizes the impurity. In other words, the impurity sits at the
center of a hole in the electron gas. The similarity to the situation
for a positive impurity is emphasized in Fig.~6 where we superpose the
screening densities for $Z=1$ and $Z=-1$ at $r_s = 10$. 
The $Z=-1$ screening density is in fact very similar to the continuum
part of the $Z=1$ screening cloud. Thus, at sufficiently large $r_s$,
the H$^-$ ion for $Z=1$ effectively acts as an external $Z=-1$
impurity as far as the rest of the electron gas is
concerned. This emphasizes that the H$^-$ configuration exists as a
well-defined entity in this range of $r_s$ values.

In all cases we find that the screened $Z = -1$ potential does not
support any bound states. We mention this since it has previously 
been claimed\cite{ghazali95} 
that the introduction of a negative test charge into a 2D
gas could give rise to potentials which would bind an electron (or
rather, a second negative test charge). To make contact with this
earlier work we compare in Fig.~7 the nonlinearly screened potentials
with those obtained on the basis of linear response theory. One curve
shows the screened potential as obtained in the RPA. This potential is
defined in Fourier space as
\beq
v_{sc}^{RPA}(q) = {v(q) \over \epsilon^{RPA}(q)} = {2\pi/q \over
\epsilon^{RPA}(q)}\,,
\eeq 
where the RPA dielectric function is given by\cite{footnote1}
\beq
\epsilon^{RPA}(q) = 1+v(q) \chi_0(q)\,.
\label{epsRPA}
\eeq
At this level of approximation, only Hartree interactions are relevant
and the screened potential defined above is the potential experienced by
a $Z=-1$ test charge  moving in the vicinity of the impurity. 
A second curve shows
the linearly screened potential, $v_{sc}^{LFC} =  v(q)/\epsilon^{LFC}$,
obtained when local field corrections
(LFC) are included in the definition of the dielectric function. Within
the LDA, this is given by
\beq
{1 \over\epsilon^{LFC}(q)} = 1-{v(q) \chi_0(q) \over 1 + (v(q) + 
v_{xc}')\chi_0(q)}\,,
\eeq
where $v_{xc}'(n_0) = dv_{xc}(n_0)/dn_0$ is the LDA local
field correction, i.e., in standard notation, $v(q)G(q)=-v_{xc}'(n_0)$. 
The potential $v_{sc}^{LFC}$ is simply the Hartree potential obtained 
with a screening charge density that includes the effects of LFC. As can
be seen in Fig.~7 for $r_s = 4$, this potential has a large negative
region in real space and supports a bound state for a unit negative
test charge of one electron mass. The authors of Ref.~\cite{ghazali95}
then argue
that the potential also binds an {\it electron} in the 2DEG and that
this binding may be a relevant pairing mechanism.
However this conclusion is invalid for several reasons. First, 
the Hartree potential for the {\it nonlinearly} screened impurity (the
short-dashed curve in Fig.~7) has a much shallower attractive portion as
compared to the corresponding result of the linear calculation
(long-dashed curve). A similar result was found analytically for a point
charged screened by a uniformly charged disk\cite{nagy99}.
But more importantly, an electron, as
opposed to a negative test charge, also feels the effect of the induced
xc potential. With this contribution included, the full
self-consistent nonlinear potential $\Delta v_{\rm eff}$ defined in 
Eq.~(\ref{veff}) has a very shallow attractive part, and there is no 
tendency for bound state formation, as 
confirmed numerically. At a more fundamental level, the question of
pairing involves the effective interaction between pairs of electrons at
the Fermi level which typically have a high relative velocity. From this
point of view it is unclear to what extent a {\it statically} screened 
potential for a test charge can be used as an estimate of the pairing
interaction.

\subsection{$Z=1$, $d \ne 0$}
   
We next consider the effect of moving the impurity charge out of the
plane of the 2DEG. This is the situation corresponding to an external
charge incident on the 2DEG from the outside, or a remote ionized
impurity in a semiconductor heterostructure.\cite{ando82,fletcher90}

Fig.~8 shows the screening density for a few values of $d$. It can 
be seen that the amplitude of the induced density decreases rapidly 
with increasing $d$ and the density becomes more spread out along the
plane. The reason for this behaviour is that the potential loses its
Coulomb singularity as soon as the impurity is out of the plane, and 
therefore becomes smoother and weaker. In Fig.~9 we show the nonlinear
screening density for the case $r_s = 2$ and $d=2$ a.u. together 
with the bound state and continuum contributions. The bound state in 
this case is very
shallow and extended, in fact more extended than the total density
itself. This is possible since the continuum contribution to the density
has a similar extent but is of opposite sign, so that there is a
cancellation of the densities in the asymptotic region.

Also shown in Fig.~9 is the screening density as determined by linear
response theory. Despite the fact that the nonlinear screening response
has a bound state contribution, the linear response result is seen to 
agree very well. The reason for this agreement is not obvious and will 
be explained in detail in Sec.~IV.
In addition, the
density is already fairly close to the classical image theory
result $(d/2\pi)(r^2+d^2)^{-3/2}$. This expression for $d = 2a_0$
gives a density at the origin of 0.04 a.u. as compared to the nonlinear
result of 0.035 a.u. With further increases in $d$ the nonlinear density 
rapidly approaches the classical screening charge density.

\subsection{Stopping Power}

As discussed in the Introduction, there have been a number of
calculations of 2D stopping powers. Our purpose here is to present
results which are based on the full nonlinear screening charge densities
and potentials. Within the so-called kinetic theory framework, the 
stopping power at low velocities for a projectile
moving with velocity $v$ in the plane of the 2DEG is given by the
expression\cite{bonig89}
\beq
S = n_0 v v_F \sigma_{tr}(E_F)\,,
\eeq
where $\sigma_{tr}(E_F)$ is
the momentum-transfer cross section defined in terms of the scattering
phase shifts by\cite{stern67}
\beq
\sigma_{tr}(E_F) = {4\over v_F} \sum_{m=0}^\infty \sin^2[\eta_m(E_F)-
\eta_{m+1}(E_F)]\,.
\eeq
To leading order in the velocity, it is sufficient to determine the
scattering phase shifts using the static nonlinearly screened potentials
calculated in the present paper.

In Fig.~10 we show the stopping power as a function of the projectile
charge $Z$ for $d=0$ and $r_s = 2$.
For small $Z$, $S$ has the expansion $S = S_1 Z^2 + S_2 Z^3+\dots$
where the first two terms are the linear and quadratic response results,
respectively. To emphasize the deviations from linear response, we
present the results in the form $S/(vZ^2)$. In this representation,
the stopping power including the quadratic response correction appears
as a straight line with slope $S_2/v$. This correction was previously
calculated within quadratic  RPA\cite{bergara99} and is shown in Fig.~10
as the straight line. The fact that the line is tangent to the nonlinear
Hartree curve at $Z = 0$ shows that the dynamic quadratic response
formulation is consistent with the kinetic theory approach for small
values of $Z$ and for low velocities. However, the validity of the 
quadratic response result
is mainly limited to negative charges down to $Z \simeq -1$ and breaks
down completely for $Z \simeq 1$. It is thus apparent that
nonlinearities are generally very important. Furthermore, 
it can be seen that the
inclusion of xc leads to a large enhancement of the stopping power in
the range $-1 \le Z \le 1$, even to a greater extent than found in 
3D.\cite{echenique91} This emphasizes that properties associated with
the scattering of continuum
states will be strongly affected by the xc part of the self-consistent 
potential.

The behaviour of the stopping power as a function of density is also of
interest and in Fig.~11 we show $r_s S/v$ {\it vs.} $r_s$ for $Z =1$. 
The quantity $S/v$ has the physical interpretation of a friction
coefficient\cite{bonig89} and is actually a monotonically decreasing
function of $r_s$ with a finite limiting value of $\pi Z^2$ for
$r_s \to 0$. This result follows\cite{nagy95} from the 2D 
momentum-transfer cross section 
$\sigma_{tr}(E_F) = Z(2\pi/v_F^2)\tanh(\pi Z/v_F)$ which
is obtained when scattering from a bare Coulomb potential, $-Z/r$, is
considered. This is the exact result in the high density limit and 
simply reflects the fact that screening is not important for high
energy scattering. As a result of this finite limiting value, the
quantity $r_sS$ plotted in Fig.~11 shows a maximum at $r_s \simeq 1$
before going to zero for $r_s \to 0$.

The results obtained previously by Wang and Ma\cite{wang97} for $Z =1$
are qualitatively similar to our results although there are some
quantitative differences. We recall that they approximated the screened
potential by the linearized TF potential and adjusted a screening 
parameter in order to satisfy the FSR. This potential is everywhere
attractive and misses the repulsive region shown in Fig.~1 which is
associated with H$^-$ formation at low densities. These differences
become more apparent when the stopping powers for $Z =1$ and $Z=-1$
are compared. This is shown in the inset to Fig.~11 where the ratio of
these stopping powers is given. The ratio tends to unity for $r_s \to 0$
since linear response is valid in this limit. However, the ratio also
tends to unity for {\it large} $r_s$ since as we have seen, the H$^-$
configuration is effectively the same as an external negative charge
insofar as the scattering of continuum states is concerned 
(see Fig.~6).  In contrast,
for an assumed model potential, Wang and Ma find the ratio of stopping
powers to be close to 2 for all $r_s$. This points to the danger of
using model potentials which do not properly account for the true nature
of the electronic screening.
Another difficulty of the model potential approach is that
differences associated with different screening approximations are no
longer apparent. This is emphasized in Fig.~11 by the plot of the
stopping power in the Hartree approximation. The results are
qualitatively similar to those with xc, but there are important 
differences.  Clearly additional input is needed in a model potential 
approach to capture these differences.

Finally, we compare with the results of Krakovsky and
Percus\cite{krakovsky95} for $Z =-1$. As mentioned in the
Introduction, they obtain the screening charge density using the
Sj\"olander-Stott integral equation\cite{sjolander72} which is a 
different way of
introducing exhange and correlation. This induced density is shown in a
second paper\cite{krakovsky95b} and is qualitatively similar to our 
results in Fig.~5. They then use this density to construct an
effective scattering potential for which the scattering phase shifts are
determined. However they give no details about how their potential is
constructed nor what interaction effects, such as exchange and
correlation, are included. Nevertheless, the stopping powers they obtain
for $Z = -1$ in 2D are similar to our results. We find in fact that
our Hartree and xc stopping powers bracket their results with deviations
which are typically less than 20\%.  

\section{High Density Limit}

We briefly commented earlier on the behaviour of the screening cloud in
the high density limit ($E_F \to \infty$, $r_s \to 0$). In this Section
we investigate this question in more detail and in particular, establish
the connection between linear and nonlinear screening for a 2DEG.
This problem in 3D was touched on in an early paper by 
Butler\cite{butler62}
which dealt with the possibility of positronium formation in metals.
Using a square well model potential, it was shown
that the total electron density at the origin is an analytic function of
the strength of the potential. This in fact is a special case of a more
general result obtained by Kohn and Majumdar\cite{kohn65} concerning the
analyticity of physical properties of the entire electron gas.
Butler then used his result regarding the analyticity of the density
to argue that the total density at the 
position of a positron in a metal could be obtained by perturbation 
theory (to all orders) even if the screened positron potential had a 
bound state. This assertion was confirmed for the specific 
example of the square well model where lowest order perturbation theory
already provided a good estimate of the exact total density in the 
limit of high electron gas densities.

We shall now prove a theorem which demonstrates that some of Butler's
conclusions have a much broader range of validity. It is unnecessary 
to specify either the form of the externally imposed potential nor the
dimensionality of the system.
The physical situation we consider is a {\it noninteracting}
Fermi gas in the presence of an external potential $\lambda V(\bfr)$. 
The effect of electron-electron interactions will be dealt with later. 
The potential is
assumed to be smooth and to approach zero sufficiently rapidly as 
$|\bfr| \to \infty$. The parameter $\lambda$ is an arbitrary coupling 
constant whose physical value is unity. 

The problem is to determine the induced density $\Delta n(\bfr)$ which
is due to the introduction of $\lambda V(\bfr)$. This quantity is
conveniently defined in terms of the Green's operator
\beq
\hat G(z) = {1\over z-H} = {1\over z-H_0 -\lambda V}
\label{1}
\eeq
where $H_0$ is the unperturbed hamiltonian. In a spatial representation,
the Green's function is given by
\bea
G(\bfr,\bfr',z)& = &\langle \bfr | \hat G(z) | \bfr' \rangle
\nonumber\\
&=& \sum_i {\psi_i(\bfr) \psi_i^*(\bfr') \over z - E_i}
\label{2}
\eea
where $\psi_i(\bfr)$ is an energy eigenstate of $H$. From this
we see that
\beq
\sum_i |\psi_i(\bfr)|^2 \delta (E - E_i) = -{1\over \pi}
\mbox {Im} G(\bfr,\bfr,E+i\epsilon)\,.
\label{3}
\eeq
Thus the total density is given by
\beq
n(\bfr) = \sum_{i(occ)}  |\psi_i(\bfr)|^2 = -{1\over \pi}
\int_{-\infty}^{E_F} dE\, \mbox {Im} G(\bfr,\bfr,E+i\epsilon)\,,
\label{4}
\eeq
where the upper limit of the energy integration is the Fermi
energy $E_F$ below which all states are occupied.

The change in density from the unperturbed situation is clearly
\bea
\Delta n(\bfr) &=& n(\bfr)-n_0(\bfr) \nonumber \\
&=& -{1\over \pi}\int_{-\infty}^{E_F} dE\, \mbox {Im} \left [ 
G(\bfr,\bfr,E+i\epsilon) - G_0(\bfr,\bfr,E+i\epsilon)\right ] \,,
\label{5}
\eea
where $G_0(\bfr,\bfr´,z)$ is the Green's function for $\lambda =0$.
It is the screening density $\Delta n(\bfr)$ that we wish to 
investigate in the high density limit.
Since the potential is multiplied by the coupling constant
$\lambda$, the density as given by Eq.~(\ref{5}) depends
implicitly on this parameter. The variation of the induced
density with $\lambda$, holding $E_F$ fixed, is simply
\beq
{\partial \Delta n(\bfr; \lambda) \over \partial \lambda} =
-{1\over \pi} \int_{-\infty}^{E_F} \mbox {Im} \langle \bfr |
{\partial \hat G(E+i\epsilon) \over \partial \lambda} | \bfr \rangle
\,.
\label{6}
\eeq
By making use of the Dyson equation
\beq 
\hat G(z,\lambda+\delta \lambda) = \hat G(z,\lambda)+
\hat G(z,\lambda)\delta \lambda V \hat G(z,\lambda+\delta \lambda)\,,
\label{7}
\eeq
we find
\beq
{\partial \hat G(z,\lambda) \over \partial \lambda} = \hat G(z,\lambda)
 V \hat G(z,\lambda)\,.
\label{8}
\eeq
Thus, the variation of the induced density can be expressed in the form
\beq
{\partial \Delta n(\bfr; \lambda) \over \partial \lambda} =
-{1\over \pi} \int d\bfr' V(\bfr') 
\int_{-\infty}^{E_F} dE\,\mbox {Im} \left [G(\bfr,\bfr',E+i\epsilon)
G(\bfr',\bfr,E+i\epsilon)\right ]\,.
\label{9}
\eeq
We note at this point that this quantity still depends implicitly on
$\lambda$ through the full Green's function $G(\bfr,\bfr',z)$.

We now make use of the spectral representation of the Green's function
in Eq.~(\ref{2}).
The product $\psi_i(\bfr) \psi_i^*(\bfr')$ can be treated as real
since a complex wave function at the energy $E_i$ must have a
time-reversed pair which is its complex conjugate. Thus, we have
\bea
{\partial \Delta n(\bfr; \lambda) \over \partial \lambda} &=&
\int d\bfr'\, V(\bfr')
\sum_{ij} \psi_i(\bfr) \psi_i^*(\bfr') \psi_j(\bfr')
\psi_j^*(\bfr) \nonumber \\
&\times &P{1 \over E_i -E_j}
\int_{-\infty}^{E_F} dE \left[ \delta(E-E_i)-
\delta(E-E_j)\right ]\,.
\label{10}
\eea
This expression is formally exact and is the crucial step in the
argument needed to arrive at the final conclusion. It is clear that the
only terms which give a nonvanishing contribution are those with 
$E_i < E_F$ and $E_j > E_F$, or vice-versa. It is these terms which 
give a finite result in the $E_F \to \infty$ limit. If this limit is 
taken first for a given $(i,j)$, the result is {\it zero}.
In other words, it is important to sum
over all states before the $E_F \to \infty$ limit is taken.
We also note that the $i=j$ terms in
Eq.~(\ref{10}) do not present any difficulty despite the appearance of
the energy denominator.
From Eq.~(\ref{9}), these terms contribute
\bea
&&\sum_{i} |\psi_i(\bfr)|^2 |\psi_i(\bfr')|^2 \mbox {Im} \lim_{\epsilon \to
0} \int_{-\infty}^{E_F} dE\,  {1 \over (E + i\epsilon - E_i)^2}
\nonumber \\
&&\qquad  =
- \sum_i |\psi_i(\bfr)|^2 |\psi_i(\bfr')|^2 \mbox {Im} \lim_{\epsilon \to
0} \int_{-\infty}^{E_F} dE\, {\partial \over \partial E}
{1 \over E + i\epsilon - E_i}
\nonumber \\
&&\qquad =
\pi \sum_i |\psi_i(\bfr)|^2 |\psi_i(\bfr')|^2 
\delta (E_F - E_i)\,,
\label{10a}
\eea
which is well-defined.

The energy integral in Eq.~(\ref{10}) can be written alternatively as
\beq
\int_{-\infty}^{E_F} dE \left[ \delta(E-E_i)-
\delta(E-E_j)\right ] =
 -\int_{E_F}^{\infty} dE \left[ \delta(E-E_i)-
\delta(E-E_j)\right ]\,.
\label{11}
\eeq
Using this in Eq.~(\ref{10}), we find that Eq.~(\ref{9}) is
equivalent to
\beq
{\partial \Delta n(\bfr; \lambda) \over \partial \lambda} =
{1\over \pi} \int d\bfr' V(\bfr') \int_{E_F}^{\infty} dE\, \mbox
{Im}
\left [ G(\bfr,\bfr',E+i\epsilon) G(\bfr',\bfr,E+i\epsilon)
\right ] \,.
\label{12}
\eeq
This is a more useful form for taking the
$E_F \to \infty$ limit since now the energy argument of
the Green's function is large. If $E_F \gg |V(\bfr)|$ everywhere
(we assume here a bounded potential\cite{footnote2}), the effect of the
potential on the propagation of the electron
can be neglected and the Green's function $G$ can be
replaced by the free-particle propagator $G_0$. We then have
\bea
{\partial \Delta n(\bfr; \lambda) \over \partial \lambda} &=&
{1\over \pi} \int d\bfr' V(\bfr') 
\lim_{E_F \to \infty} \int_{E_F}^{\infty} dE\, \mbox {Im} \left [
G(\bfr,\bfr',E+i\epsilon) G(\bfr',\bfr,E+i\epsilon)\right ]
\nonumber \\ &\simeq&
{1\over \pi} \int d\bfr' V(\bfr') 
\lim_{E_F \to \infty} \int_{E_F}^{\infty} dE\, \mbox {Im} \left [
G_0(\bfr,\bfr',E+i\epsilon) G_0(\bfr',\bfr,E+i\epsilon)\right ]
\nonumber \\ &=&
-{1\over \pi} \int d\bfr' V(\bfr') 
\lim_{E_F \to \infty} \int_{-\infty}^{E_F} dE\, \mbox {Im} \left [
G_0(\bfr,\bfr',E+i\epsilon) G_0(\bfr',\bfr,E+i\epsilon)\right ]\,,
\label{13}
\eea
where we have simply reversed the earlier argument to 
change the limits of integration in the last line.
Several comments are in order:
\begin{enumerate}
\item
The replacement of $G$ by $G_0$ is only valid if the energy argument is
large; these Green's functions are certainly different
at low energies. In fact, the perturbed and unperturbed systems
have completely different state spectra and $H$ may contain
bound states which do not appear in $H_0$. Nevertheless, since
Eq.~(\ref{9}) can be written in the equivalent form (\ref{12}), the
replacement of $G$ by $G_0$ in Eq.~(\ref{13}) is justified in the
$E_F \to \infty$ limit.
\item The final line in Eq.~(\ref{13})
is identical to the result obtained by linear response theory. 
Since it is {\it independent} of $\lambda$, it
can be integrated from $\lambda = 0$ to  $\lambda = 1$
to give 
\beq
\Delta n(\bfr) = 
-{1\over \pi} \int d\bfr' V(\bfr') 
\lim_{E_F \to \infty} \int_{-\infty}^{E_F} dE\, \mbox {Im} \left [
G_0(\bfr,\bfr',E+i\epsilon) G_0(\bfr',\bfr,E+i\epsilon)\right ]\,.
\label{14}
\eeq
\item The final result in Eq.~(\ref{14}) does not depend on the assumed
dimensionality and is valid even when the potential $V(\bfr)$ supports
bound states. These bound states of course do not appear in the spectrum
of $G_0$ but are nevertheless included in $\Delta n(\br)$. The result 
given in Eq.~(\ref{14}) will be referred to as the {\it high density 
screening theorem}.
\end{enumerate}

As an application of Eq.~(\ref{14}) we consider an ideal uniform gas in
$D$ dimensions.
The use of the spectral representation for the Green's function
gives in this case
\bea
-{1\over \pi} \int_{-\infty}^{E_F} &dE&\, \mbox {Im} \left [
G_0(\bfr,\bfr',E+i\epsilon)
G_0(\bfr',\bfr,E+i\epsilon)\right ] \nonumber \\
&=& {2\over L^{2D}}
\sum_{\bfk\bfk'} e^{i(\bfk'-\bfk)\cdot(\bfr-\bfr')} P{1\over
E_{\bfk'}-E_\bfk} \left [ f(E_{\bfk'}) - f(E_\bfk)\right ]
\nonumber \\
&=& -{1\over L^D} \sum_{\bfq} e^{i\bfq \cdot (\bfr-\bfr')}
\chi_0(\bfq)
\label{15}
\eea
where the factor of 2 is for spin, $L$ is the size of the sample in a
given direction, $f(E)$ is the zero-temperature 
Fermi function and $\chi_0(\bfq)$ is the static
free particle density response function
\beq
\chi_0(\bfq) = {2\over L^D} P\sum_{\bfk} {f(E_{\bfk+\bfq})-f(E_{\bfk})
\over E_\bfk -  E_{\bfk+\bfq}}\,.
\label{16}
\eeq
Substitution of Eq.~(\ref{15}) into Eq.~(\ref{14}) gives
\beq
\Delta n(\bfr) =  -\lim_{E_F \to \infty}{1\over L^D} \sum_{\bfq}
\chi_0(\bfq) V(\bfq) e^{i\bfq \cdot \bfr}\,.
\label{17}
\eeq
This is a rigorous statement that linear response gives the correct
induced density in the high density limit in any dimension.

The static response functions in the various dimensions are given by 
the expressions
\bea
\chi_0^{1D}(q) &=& {2\over \pi q} \ln { {2k_F + q} \overwithdelims
|| {2k_F - q}}\,,\nonumber \\
\chi_0^{2D}(q) &=& {1\over \pi} \theta (2k_F -q) + {1\over \pi}
\left [ 1-\sqrt {1- ( 2 k_F / q )^2 } \right ] 
\theta(q-2k_F)\,,\\
\chi_0^{3D}(q) &=&{k_F\over 2\pi^2}\left[ 1 - {k_F \over q}\left (
1 - {q^2 \over 4 k_F^2} \right ) \ln { {2k_F - q} \overwithdelims
|| {2k_F + q}} \right ]\nonumber \,.
\eea
In each case, the $q\to 0$ limit of $\chi_0(q)$ is the density of states
$g(E_F)$ at the Fermi level
\bea 
\chi_0^{1D}(0) &=& {2\over \pi k_F}\,,\nonumber \\
\chi_0^{2D}(0) &=& {1\over \pi}\,,\\ 
\chi_0^{3D}(q) &=&{k_F\over \pi^2}\nonumber\,.
\eea
If the potential is smooth we can suppose that $V({\bf q})$ is
effectively zero for $q > q_{\rm max}$ with $q_{\rm max}  \ll k_F$. 
In this case, Eq.~(\ref{17}) reduces to
\beq
\Delta n(\bfr) =  - \lim_{E_F \to \infty}\chi_0(0) V(\bfr)\,,
\label{22}
\eeq
which is recognized as the linearized TF approximation for the density.

It should be noted that 2D is special for a number of reasons. First,
since $E_F$ is proportional to the density $n$, the full TF 
approximation is
itself linear. Second, the free-particle response function has the 
constant value $\chi_0(\bfq) = 1/\pi$ for all wavevectors up to $2k_F$.
Thus, the replacement of $\chi_0({\bf q})$ by $\chi_0(0)$ is
in some sense a less restrictive approximation than it is in other
dimensions. In particular, if the only finite
Fourier components of $V(\bfr)$ occur for $q < 2k_F$ (which
is always the case when $k_F \to \infty$), we have
\beq
\Delta n(\bfr) =  -{1\over \pi} V(\bfr)\,.
\label{18}
\eeq
The analogous result in 3D is $\Delta n(\bfr) = 0$ 
while the 1D result does not have a finite limit. 
Thus the induced density only has a nontrivial limiting value in 2D.

The result for the 2D screening charge in Eq.~(\ref{18}) is all the more
remarkable if we note that the
potential $V(\bfr)$ can in fact support bound states, as it must in 2D
if the potential is purely attractive. It is worth verifying this
explicitly by means of a numerical example. We consider the following
model potential 
\beq
V(r) =\cases{V_0 \sin^2 \left ( {2\pi r\over r_0} \right ) \,,
\quad r\le r_0 \,, \cr
0 \,, \quad r > r_0\,, \cr}
\label{modelpot}
\eeq
which has no physical basis but is chosen simply to illustrate the
point. It is a repulsive double barrier potential for $V_0 > 0$ and an
attractive double well potential for $V_0 < 0$. 
With this potential we calculate the induced density by solving the 
Schr\"odinger equation for all states below $E_F$. 
In Fig.~12 we show results for a repulsive potential with
$V_0 = 0.125$ $H$ and $r_0 = 5$ a.u. as a function of gas density. 
The screening charge is due to continuum states in this case and it 
can be seen that by $r_s = 0.5$, the induced density is very close to 
the prediction of
Eq.~(\ref{18}). Similarly, if an attractive potential is placed in a gas
with $r_s = 0.5$ ($E_F = 4.0 $ $H$), we obtain the bound and continuum 
charge densities shown in Fig.~13. Here, the magnitude of $|V_0|$ was 
adjusted to vary the number of bound states. 
Two cases were considered: (a) $V_0 = -0.125$ $H$ for which 
a single $m=0$ bound state is found and (b) $V_0 = -0.25$ $H$,
for which two $m=0$ bound states exist. The bound and continuum
densities are quite different for these two cases, but the total density
agrees very well with the result in Eq.~(\ref{18}), even though the high
density limit has only marginally been reached. These examples confirm
very nicely the behaviour stipulated by the high density screening
theorem. 

We can now apply the theorem to explain some of the results found in 
the $Z=1$
calculations of Sec. III. Of course interactions were included there, 
but the Kohn-Sham system of equations correspond to a system of {\it
noninteracting} electrons moving in the total effective potential in
Eq.~(\ref{veff}). In Fourier space, we have 
\beq
\Delta v_{\rm eff}(q) = v_{\rm ext}(q) + v(q)\Delta n(q) + v_{xc}'(n_0)
\Delta n(q)\,.
\label{veff2}
\eeq
We have here linearized the xc potential since, in the high density
limit of interest, the induced density is small compared to the 
background
density. Identifying $\Delta v_{\rm eff}(q)$ as given by 
Eq.~(\ref{veff2}) with $V(q)$ in Eq.~(\ref{17}), 
we have
\beq
\Delta n(q) = - \chi_0(q) \Delta v_{\rm eff}(q) 
\eeq
which implies
\beq
\Delta n(q) = - {\chi_0(q) v_{\rm ext}(q) \over 1 + (v(q)
+v_{xc}'(n_0))\chi_0(q)}\,.
\eeq
This is simply the induced density obtained within linear response
theory, treating Hartree and xc interactions at the mean-field level.
An example of the linear response result was already given in Fig.~9
for the case of a $Z=1$ charge outside the plane of the 2DEG. The close
agreement of this density with the nonlinear screening density is
confirmation of the high density screening theorem.

For high densities, $v_{xc}$ is in fact dominated by the exchange 
potential $v_x = -C_x/r_s$ and we have $v_{xc}'(n_0)
\simeq -(\pi C_x /2)r_s$.
Thus in the limit $r_s \to 0$, the xc term can be dropped and the 
induced density is simply given by the result
\beq
\Delta n(q) = - {\chi_0(0) v_{\rm ext}(q) \over 1 + v(q) \chi_0(0)}\,.
\label{denTFq}
\eeq
The corresponding effective potential is
\beq
\Delta v_{\rm eff}(q) = { v_{\rm ext}(q) \over 1 + v(q)
\chi_0(0)}\,.
\label{vTFq}
\eeq
The real space quantities are obtained from these expressions by an
inverse Fourier transform.

We now consider the particular example of a point charge $Z$ in the
plane of the gas. With $v(q) = 2\pi/q$ in 2D, Eq.(\ref{vTFq}) 
becomes
\beq
\Delta v_{\rm eff}(q) = -{2\pi Z \over q+2}\,,
\label{27}
\eeq
and in real space we obtain the TF potential\cite{stern67}
\bea
v_{\rm TF}(r) &\equiv& -Z \int_0^{\infty} dq\, {q\over q+2}
J_0(qr)\nonumber \\
&=& -Z \int_0^\infty dx\, {x e^{-2x}\over (x^2 + r^2)^{3/2}}\,.
\label{28}
\eea
It is a monotonic potential which behaves as $-Z/r$ for $r \to 0$ and 
falls off as $-Z/4r^3$ for $r \to \infty$. Being negative definite, it
will support at least one bound state for any value of $Z$.
For $Z= 1$ the bound state energy is found to be $E_0 = - 0.2862$ $H$
which agrees with the value found by Stern and Howard\cite{stern67}.
This is the limiting value of the bound state eigenvalue shown in
Fig.~3.

We thus come to the conclusion that the $r_s \to 0$ limit of the
Kohn-Sham effective potential is the TF potential in Eq.~(\ref{28}). To
illustrate this we show in Fig.~14 the induced density for $Z=1$ 
and a very high density corresponding to $r_s = 0.2$. Also shown are 
the bound and continuum components which add to the total. The latter is
also displayed as $r\Delta n(r)$ in Fig.~15 together with the TF
density as obtained from Eq.~(\ref{denTFq}). It can be seen that the DFT
density simply oscillates about this limiting value. With decreasing
$r_s$, the amplitude of these oscillations decrease and the DFT
density uniformly approaches the TF density.

For completeness, we note that the result corresponding to
Eq.~(\ref{27}) in 3D is
\beq
\Delta v_{\rm eff}(q) = -{4\pi Z \over q^2+q_{TF}^2}\,,
\label{29}
\eeq
with $q_{TF} = \sqrt{4k_F/\pi}$ the Thomas-Fermi wavevector.
The real-space potential in this case is
\beq
v_{\rm eff}(r) = - {Z e^{-q_{TF} r} \over r}\,.
\eeq
With increasing $k_F$, the range of the potential decreases. Thus at
some point the potential will no longer support a
bound state, and all states in the presence of the external
charge will be continuum scattering states. As alluded to earlier, the
screening of the external potential becomes more effective with
increasing density.
This is in marked contrast to the situation in 2D where the
screened potential has a density independent limit which continues to
support bound states in the high density limit.

In summary, if a potential $V(\bfr)$ is introduced into a 2D gas
and the induced density is constructed by summing up over {\it all}
eigenstates of the hamiltonian $H$ (including all possible bound
states), the induced density will be given exactly by
Eq.~(\ref{18}) in the high density limit. Remarkably, this density is 
the same as what one
would obtain by treating the potential as a weak perturbation and
applying perturbation theory.

\section{Conclusions}

In this paper we have provided a detailed discussion of the nonlinear
screening of an external point charge by a two-dimensional electron
gas. Nonlinear effects are most evident when the charge lies in the
plane of the 2DEG but rapidly decrease in importance as the charge is
moved out of the plane. As an application of the theory we have
considered the problem of 2D stopping power and have compared our
results with earlier work. Quantitative differences point to the
importance of performing self-consistent nonlinear screening
calculations.

We also derive what we refer to as the high density screening theorem.
In 2D, the theorem has the interesting consequence that the screened
potential has a density independent limiting form in the high density
limit and that the screening density is directly proportional to this
potential. Explicit calculations for model potentials confirm these
conclusions and demonstrate the surprising fact that linear response
theory is valid even when the potential acting on the gas
supports bound states. These nonintuitive results 
highlight some of the peculiarities of
electronic screening in two dimensions.

\acknowledgments
The work of I.N. has been supported by the OTKA (Grant Nos.
T034363 and T046868), and that of E.Z. by a grant from the Natural 
Sciences and Engineering Research Council of Canada. P.M.E. thanks the
University of the Basque Country UPV/EHU, the Basque Hezkuntza,
Unibertsitate eta Ikerketa Saila and the Spanish Ministerio de Ciencia y 
Tecnolog\'ia for support.

\appendix
\section{Evaluation of Hartree Potential}

In this Appendix we summarize the methods we have used to determine the
electrostatic potential of the screened impurity charge. Due to the
long-range nature of the Coulomb interaction, this is the one nontrivial
part of the calculation of the total self-consistent potential.

A charge $Z$ at the position $\bR = d\hat{\bf z}$ above the plane of
the 2DEG ($z=0$) is represented by the external charge density
\begin{equation}
n_{\rm ext}(\br) = Z \delta(\br - \bR)
\end{equation}
which gives rise to the electrostatic potential
\begin{equation}
\phi_{\rm ext}(\rho) = {Z \over (\rho^2 + d^2)^{1/2}}
\end{equation}
in the plane of the gas.
Here, $\rho$ is the radial distance from the origin.
This potential induces the electronic screening density 
$\Delta n(\br) = \Delta n(\rho)\delta(z)$ which in turn gives rise to 
an electrostatic potential that is determined by Poisson's equation
\begin{equation}
\nabla^2 \Delta \phi(\br) = 4\pi \Delta n(\br)\,.
\label{Poisson}
\end{equation}
Although this equation determines $\Delta\phi$ in three-dimensional 
space, only the behaviour within the plane of the gas is required in
Eq.~(\ref{SE}).

To solve Poisson's equation, we make use of the 2D Fourier transform 
pair:
\begin{eqnarray}
\Delta \tilde \phi(q,z) &=& \int d^2\rho \, e^{-i\bq\cdot \brho} 
\Delta \phi(\rho,z)
\nonumber \\
\Delta \phi(\rho,z) &=& \int {d^2q \over (2\pi)^2} e^{i\bq\cdot \brho}
\Delta \tilde \phi(q,z)\,.
\end{eqnarray}
Taking the Fourier transform of Eq.~(\ref{Poisson}), we obtain
\begin{equation}
{d^2 \Delta \tilde \phi \over dz^2} - q^2 \Delta \tilde \phi = 
4\pi \Delta \tilde n\delta(z)\,,
\end{equation}
which has the solution
\begin{equation}
\Delta \tilde \phi (q,z)  = \Delta \tilde\phi(q, z=0) e^{-q|z|}
\end{equation}
with
\begin{equation}
\Delta \tilde \phi (q,z=0)  = - {2\pi \over q} \Delta \tilde n\,.
\label{ftn}
\end{equation}
The screening potential in the plane of the gas is then obtained by
taking the inverse 2D Fourier transform of this equation.

Two problems arise in the numerical implementation of this procedure.
First, Eq.~(\ref{ftn}) has a $q=0$ singularity associated with the
long-range nature of the Coulomb potential, and second, a
straightforward iteration of the Kohn-Sham equations does not converge.
To deal with the first problem, we introduce an auxiliary charge density
$n_{\rm aux}(\rho)$ which gives rise to the auxiliary electrostatic
potential
\begin{equation}
\phi_{\rm aux} (\rho)  = {Z \over (\rho^2 + d_0^2)^{1/2}}\,,
\end{equation}
where $d_0$ is some fixed, nonzero parameter.
Evidently, this potential can be considered as being due to a point
charge at $\bR_0 = d_0\hat{\bf z}$. Alternatively, its 2D Fourier 
transform
\begin{equation}
\tilde \phi_{\rm aux} (q)  = {2\pi \over q} Z e^{-qd_0}
\end{equation}
implies that the auxiliary charge density in
the $z=0$ plane has the 2D Fourier transform
\begin{equation}
\tilde n_{\rm aux} (q)  = Z e^{-qd_0}\,.
\end{equation}
The corresponding real space density is therefore
\begin{equation}
n_{\rm aux} (\rho) = {Z\over 2\pi} {d_0 \over (\rho^2 + d_0^2)^{3/2}}\,.
\end{equation}
This density has a total charge of $Z$ and falls off as $\rho^{-3}$ as
$\rho \to \infty$.

We now write the external potential as
\begin{equation}
\phi_{\rm ext}(\rho) = \phi_{\rm aux}(\rho)+\Delta \phi_{\rm ext}(\rho) 
\end{equation}
where
\begin{equation}
\Delta \phi_{\rm ext}(\rho) = Z \left [ {1\over (\rho^2 + d^2)^{1/2}} -
{1\over (\rho^2 + d_0^2)^{1/2} } \right ]\,.
\label{Deltaphi}
\end{equation}
The latter is treated explicitly in real space, while the effect of
$\phi_{\rm aux}$ is accounted for by means of the following Poisson
equation:
\begin{equation}
{d^2 \Delta \tilde \phi \over dz^2} - q^2 \Delta \tilde \phi = 
- 4\pi (\tilde n_{\rm aux} - \Delta \tilde n)\delta(z)\,,
\label{Poisson2}
\end{equation}
This equation gives
\begin{equation}
\Delta \tilde \phi(q,z=0) = {2\pi \over q}(\tilde n_{\rm aux} - \Delta
\tilde n)
\end{equation}
which no longer has a $q=0$ singularity since the 
combination $\tilde n_{\rm aux} - \Delta \tilde n$ 
is charge neutral. Thus, 
its inverse Fourier transform can be calculated readily using a
fast Fourier transform (FFT) technique, and the addition of the result
to Eq.~(\ref{Deltaphi}) yields the total electrostatic potential.

The second problem concerns the convergence of the self-consistent
iterative procedure. This is dealt with by rewriting
Eq.~(\ref{Poisson2}) in the form
\begin{equation}
{d^2 \Delta \tilde \phi \over dz^2} - (q^2+\kappa^2) \Delta \tilde \phi = 
- 4\pi (\tilde n_{\rm aux} - \Delta \tilde n)\delta(z) -\kappa^2 \Delta
  \tilde \phi\,.
\end{equation}
This trivial change obviously does not change the solution of the
equation but we see that the left hand side corresponds to a {\it
screened} Coulomb interaction with screening parameter $\kappa$. 
With this equation in mind, we adopt the following iterative
procedure\cite{vanzyl99}:
\begin{equation}
\Delta \tilde \phi^{(i+1)}(q, z=0) = {2\pi \over Q}(\tilde n_{\rm aux} -
\Delta \tilde n^{(i)}) + \left ( 1 - {q\over Q} \right ) \Delta \tilde
\phi^{(i)}(q, z=0)\,,
\end{equation}
where $Q^2 = q^2 +\kappa^2$.
It is evident that when $\Delta \tilde \phi^{(i+1)}
= \Delta \tilde \phi^{(i)}$ at convergence, this procedure
yields
\begin{equation}
\Delta \tilde \phi(q, z=0) = {2\pi \over q}(\tilde n_{\rm aux} -
\Delta \tilde n)\,,
\end{equation}
which is the correct potential for the charge density $\tilde n_{\rm
aux} - \Delta \tilde n$. The use of a screened-Coulomb interaction has
the important effect of stabilizing the iterative procedure required to
achieve selfconsistency.

\begin{figure}
\centering
\scalebox{0.55}
{\includegraphics[0.5in,2.5in][8.5in,9.5in]{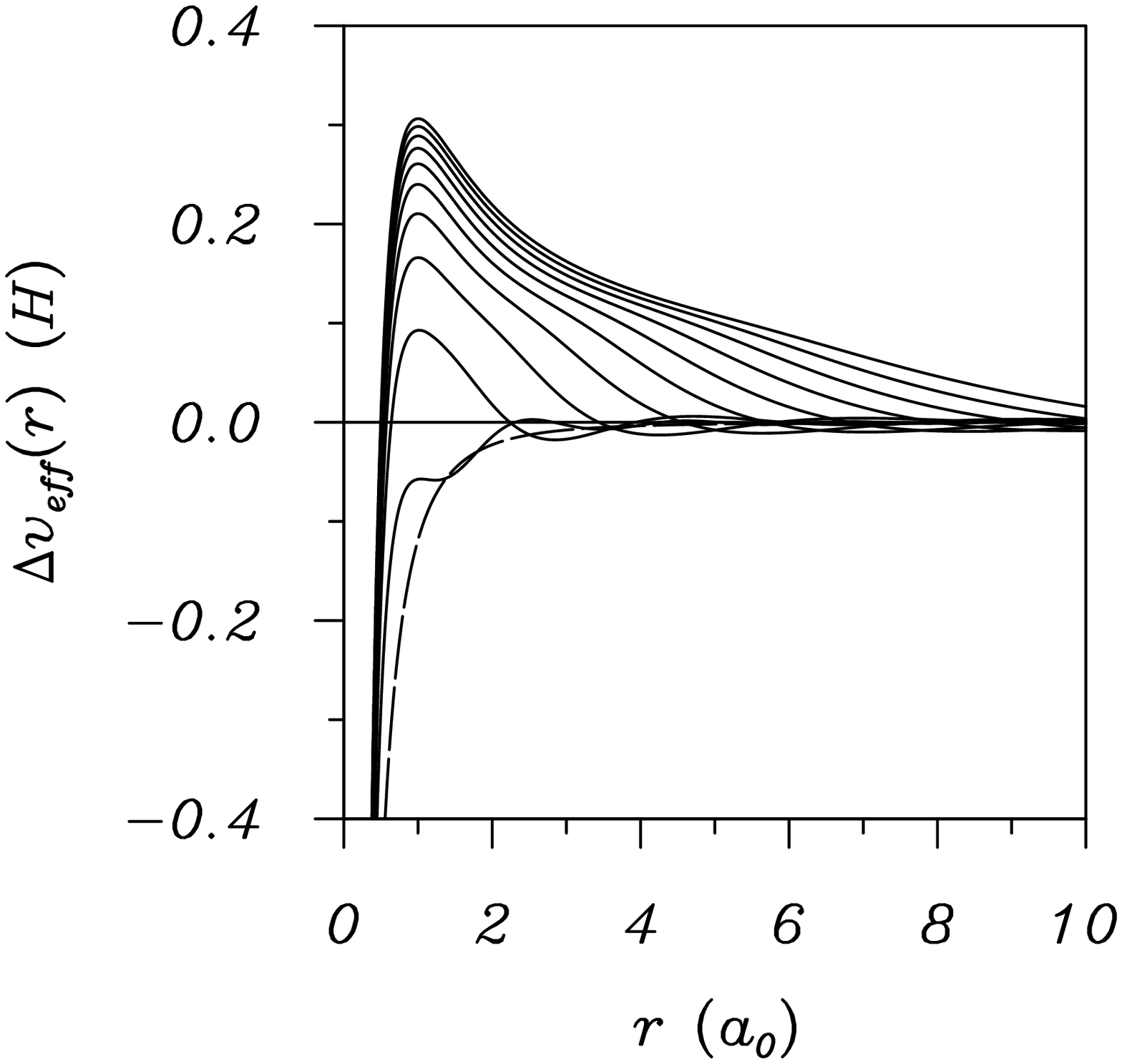}}
 \caption{Self-consistent effective potentials $\Delta v_{\rm eff}(r)$
in Hartree units as a function of position. Each solid curve
corresponds to a different $r_s$ value,
ranging from 1 for the lowest curve to 10 for the highest curve,
in steps of 1. The dashed curve is the Thomas-Fermi screened potential.
 }
 \label{fig1} 
\end{figure}

\begin{figure}
\centering
\scalebox{0.53}
{\includegraphics[0.5in,1.5in][8.5in,9.0in]{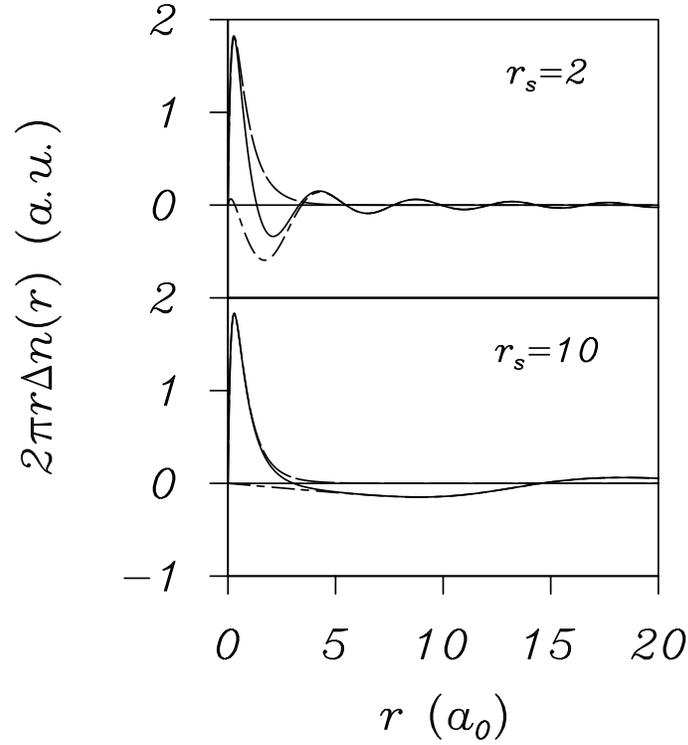}}
 \caption{Screening density as a function of position for two
$r_s$ values. The dashed curve is the bound state contribution, the
chain curve is the continuum state contribution and the solid curve is
the total.
  }
 \label{fig2} 
\end{figure}

\begin{figure}
\centering
\scalebox{0.65}
{\includegraphics[0.5in,2.0in][8.5in,8.0in]{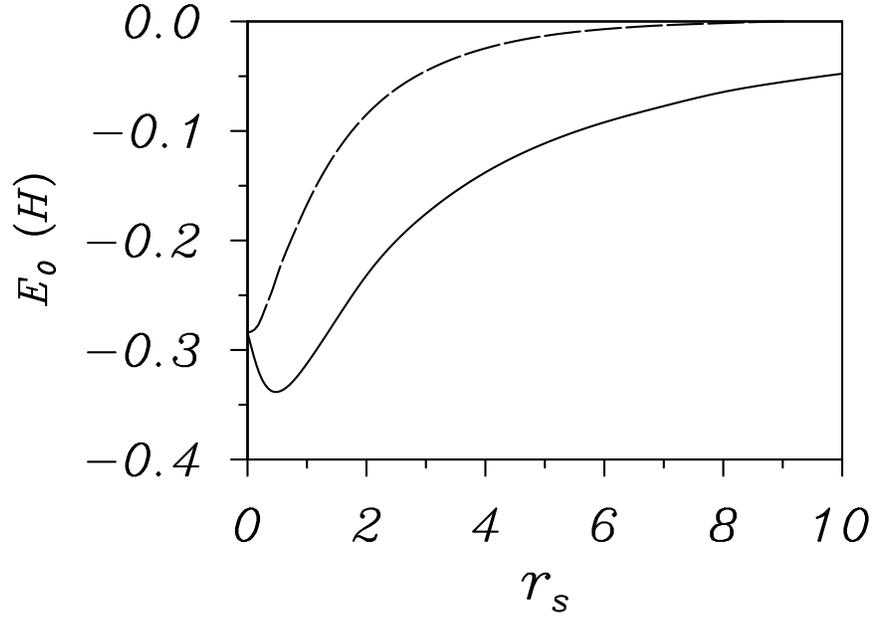}}
 \caption{Bound state eigenvalue in Hartree units {\it vs.} $r_s$. The
solid curve is for the full nonlinear DFT calculation including
exchange-correlation, while the dashed curve is the corresponding result
in the Hartree approximation.
}
 \label{fig3}
\end{figure}

\begin{figure}
\centering
\scalebox{0.6}
{\includegraphics[0.5in,1.5in][8.5in,8.0in]{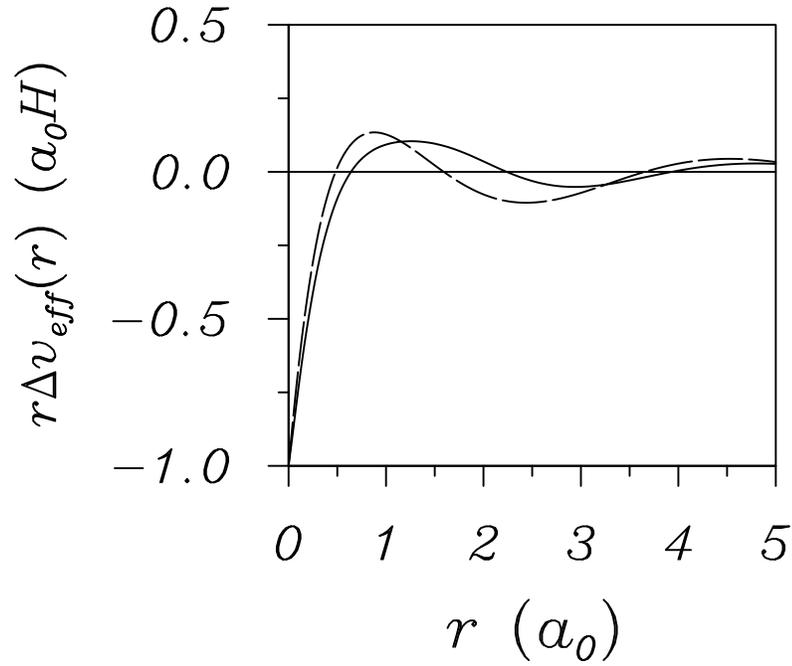}}
\caption{Self-consistent effective potentials with (solid curve) and
without (dashed curve) exchange-correlation.
 } 
\label{fig4}
\end{figure}

\begin{figure}
\centering
\scalebox{0.6}
{\includegraphics[0.5in,1.5in][8.5in,8.0in]{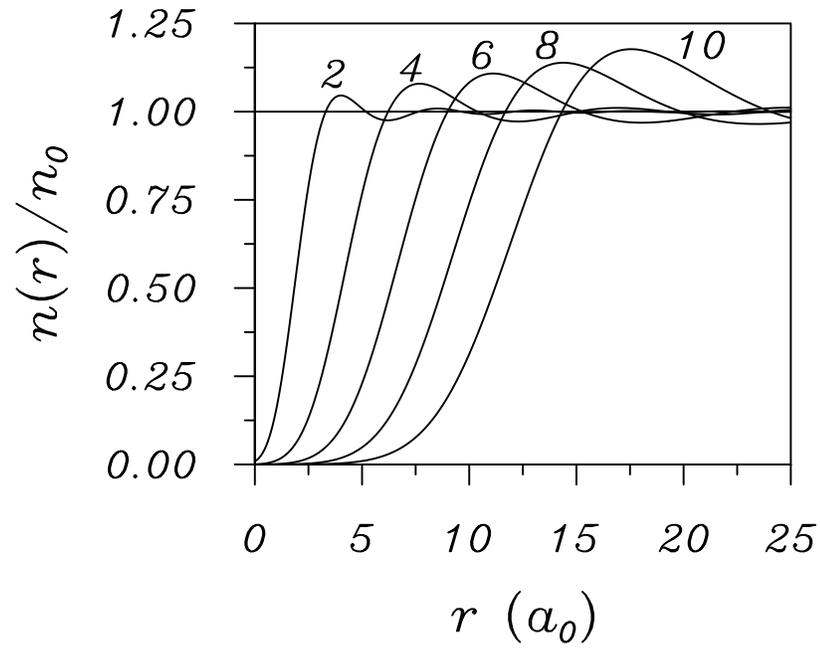}}
\caption{Normalized density as a function of position. The different
curves are labelled by the $r_s$ value.
 }
\label{fig5} 
\end{figure}
 
\begin{figure}
\centering
\scalebox{0.6}
{\includegraphics[0.5in,1.5in][8.5in,8.0in]{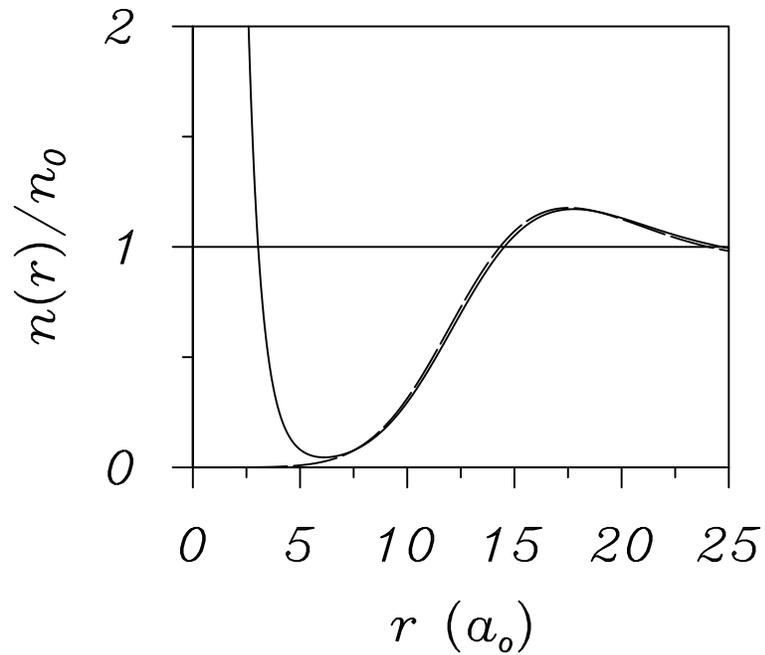}}
\caption{Normalized density as a function of position for $r_s =10$. The
solid curve is for $Z=1$ while the dashed curve is for $Z=-1$.
 }
\label{fig6}
\end{figure}

\begin{figure}
\centering
\scalebox{0.6}
{\includegraphics[0.5in,1.75in][8.5in,8.0in]{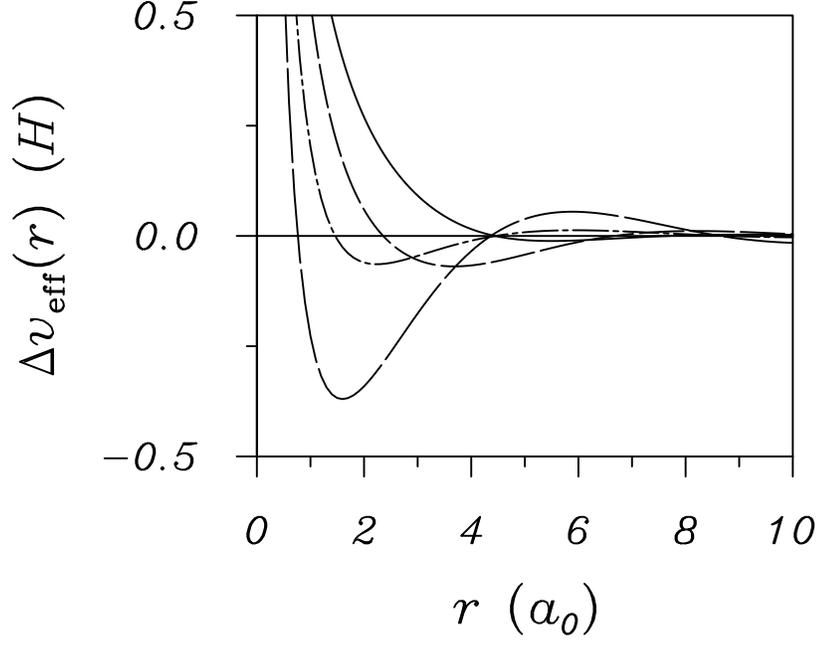}}
\caption{Comparison of nonlinearly screened and linearly screened 
potentials for $r_s =4$ and $Z=-1$. The solid curve is the nonlinearly
screened potential including exchange-correlation, $v_{\rm ext} + \Delta
v_H +\Delta v_{xc}$; the short-dashed
curve is the Hartree potential, $v_{\rm ext} + \Delta v_H$,
calculated with the same screening density as used for the solid curve ;
the chain curve is the linearly screened
potential within the RPA (i.e., the Hartree approximation); and the
long-dashed curve is the Hartree potential, $v_{\rm ext} +
\Delta v_H$, obtained with a linear screening density that includes
local-field effects (i.e., exchange-correlation).
 }
\label{fig7}
\end{figure}

\begin{figure}
\centering
\scalebox{0.6}
{\includegraphics[0.5in,2.0in][8.5in,8.5in]{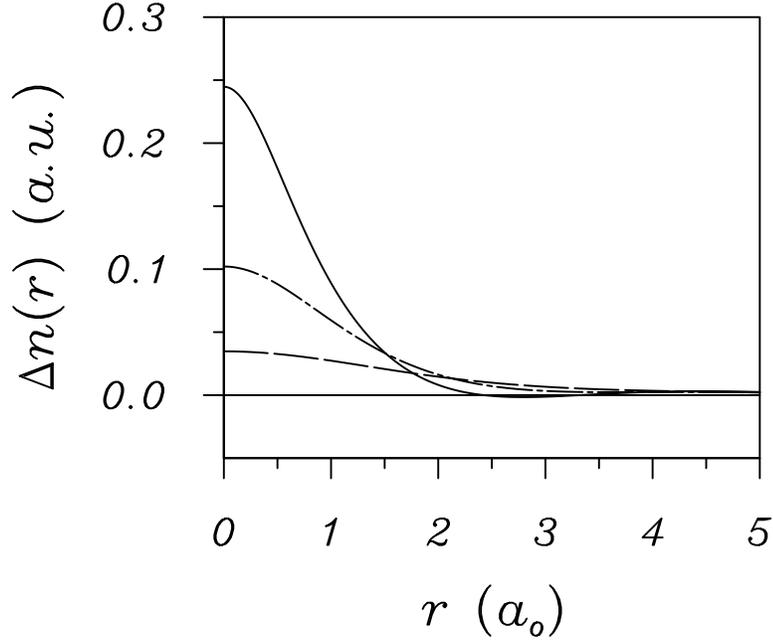}}
 \caption{Screening charge density as a function of position for
$Z=1$ and $r_s =2$ for different distances $d$ of the charge from the
plane of the electrons: $d= 0.5 a_0$ (solid curve), $d= 1.0 a_0$ 
(chain curve), $d= 2.0a_0$ ( dashed curve).
 }   
\label{fig8}
\end{figure}

\begin{figure}
\centering
\scalebox{0.6}
{\includegraphics[0.5in,1.5in][8.5in,8.0in]{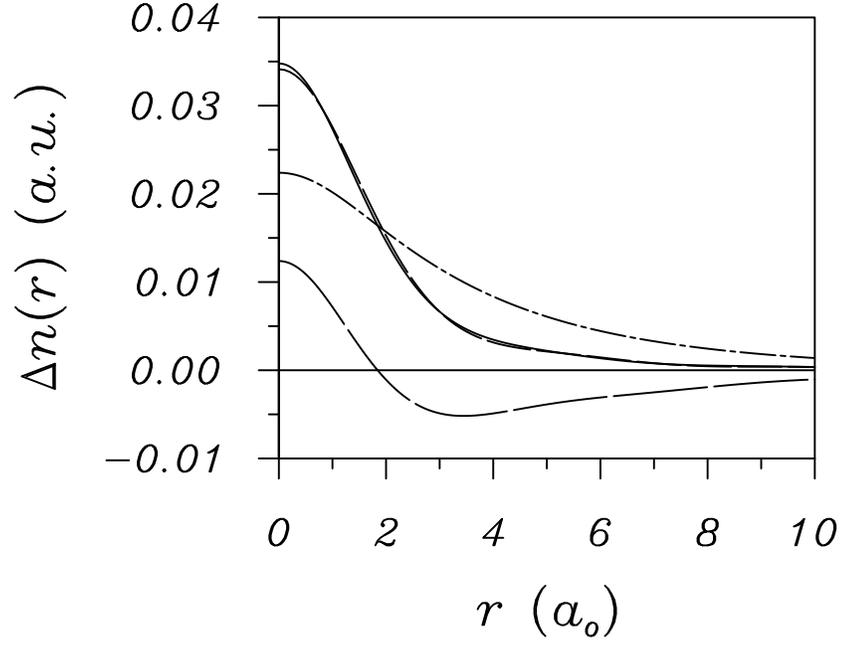}}
 \caption{Screening charge density as a function of position for $Z=1$,
$r_s =2$ and $d=2.0a_0$. The chain curve is the bound state
contribution, the long-dashed curve is the continuum contribution and
the solid curve is the total. The short-dashed curve almost coincident 
with the solid curve is the linear response result including
exchange-correlation.
 }   
\label{fig9}
\end{figure}

\begin{figure}
\centering
\scalebox{0.6}
{\includegraphics[0.5in,2.0in][8.5in,8.5in]{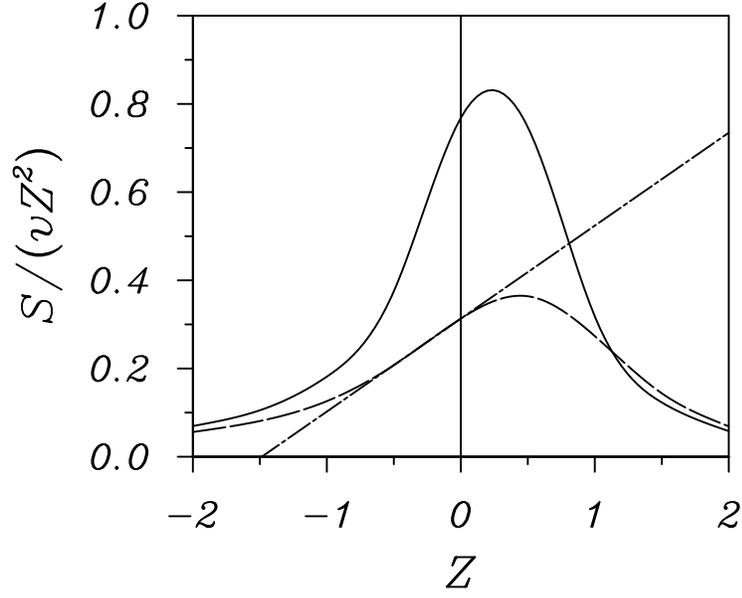}}
 \caption{Normalized stopping power as a function of the projectile
charge $Z$, for $r_s =2$ and $d=0$. The solid curve is obtained from
the nonlinear screening result including exchange-correlation; the dashed
curve shows the corresponding result in the Hartree approximation (no
exchange-correlation); the straight line (chain curve) is the quadratic
response result\protect\cite{bergara99}  $(S_1 + S_2 Z)/v$.
 }   
\label{fig10}
\end{figure}

\begin{figure}
\centering
\scalebox{0.53}
{\includegraphics[0.5in,1.5in][8.5in,8.0in]{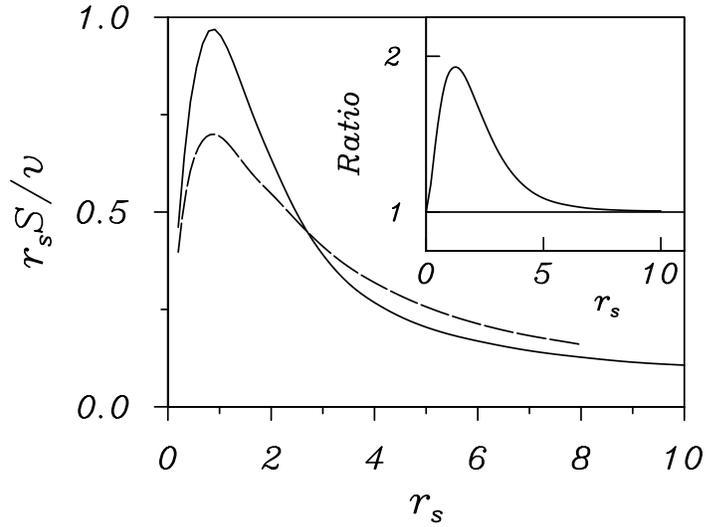}}
 \caption{Stopping power as a function of density parameter $r_s$
 for $Z=-1$. The solid curve shows the full nonlinear screening result,
including exchange-correlation, while the dashed curve gives the
corresponding result in the Hartree approximation. The inset shows the
stopping power ratio $S(Z =1)/S(Z = -1)$ as a function of $r_s$.
 }   
\label{fig11}
\end{figure}

\begin{figure}
\centering
\scalebox{0.5}
{\includegraphics[0.5in,1.5in][8.5in,10.75in]{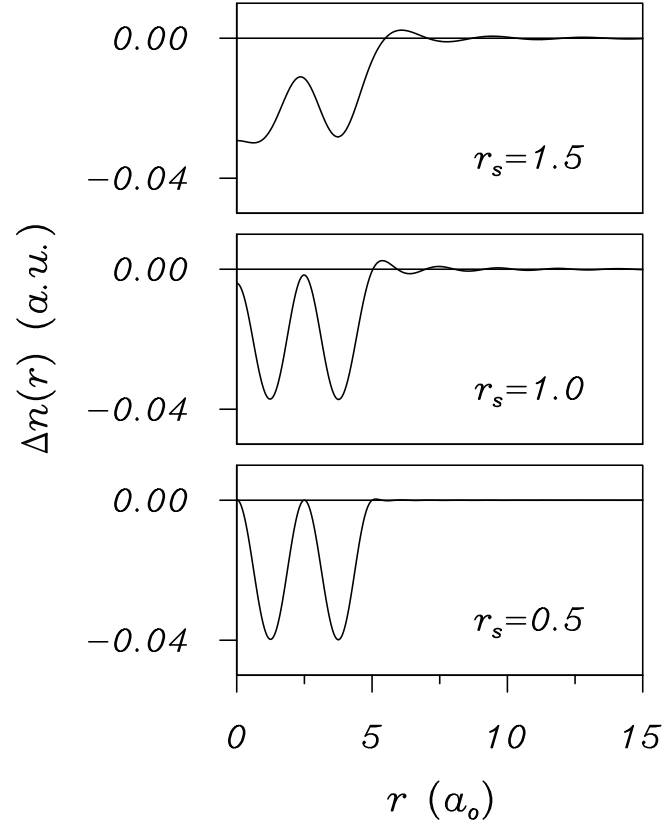}}
 \caption{Screening charge density as a function of position for the
model potential in Eq.~(\ref{modelpot}). The potential parameters are 
$V_0 = 0.125 $ H and $r_0 = 5a_0$.
 }   
\label{fig12}
\end{figure}

\begin{figure}
\centering
\scalebox{0.75}
{\includegraphics[0.5in,1.5in][8.5in,10.5in]{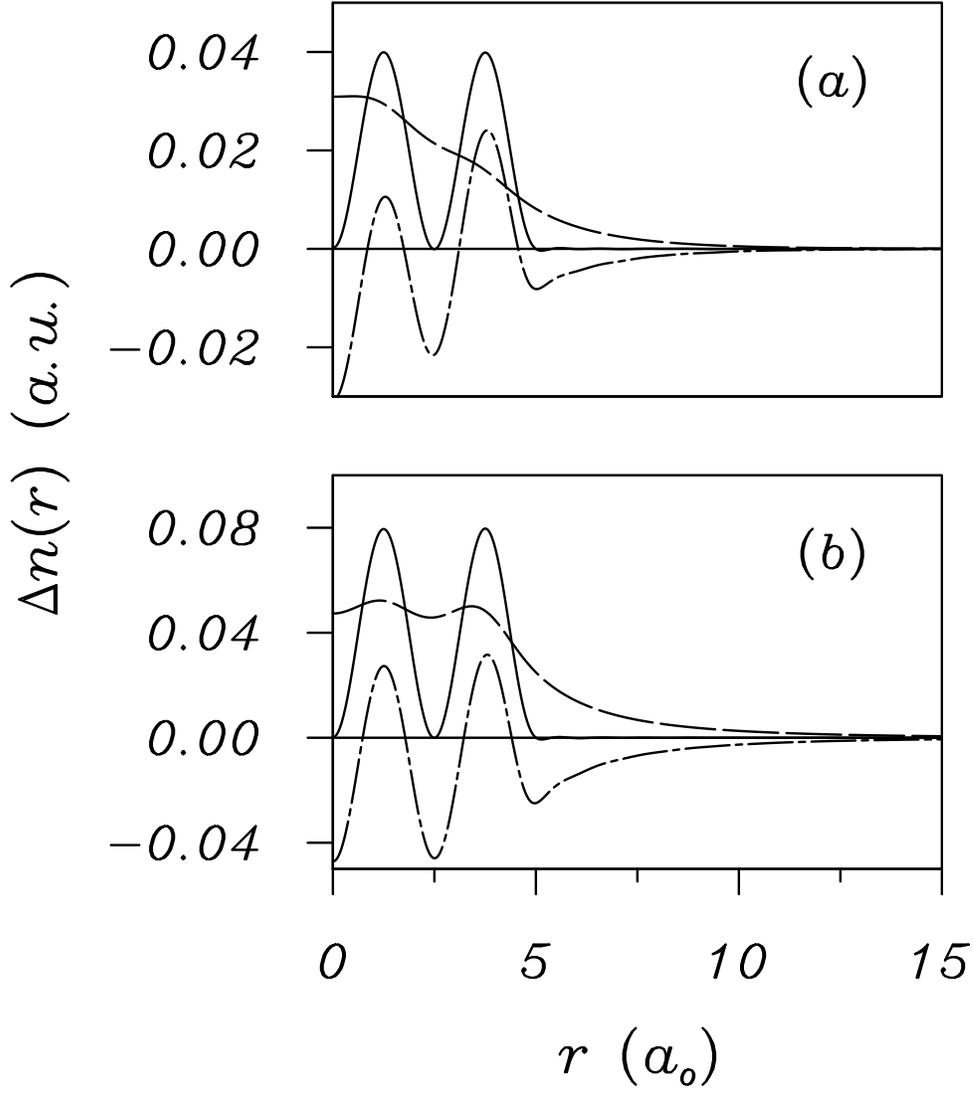}}
 \caption{Screening charge density for $r_s = 0.5$ for the model 
potenial in Eq.~(\ref{modelpot}): (a) $V_0 = -0.125\,{\rm H}$, 
$r_0 = 5\,{\rm a}_0$; (b) $V_0 = -0.25\,{\rm H}$, $r_0 = 5\,{\rm a}_0$.
The dashed curve is
the bound state contribution, the chain curve is the continuum state
contribution and the solid curve is the total.
 }   
\label{fig13}
\end{figure}

\begin{figure}
\centering
\scalebox{0.6}
{\includegraphics[0.5in,1.5in][8.5in,8.0in]{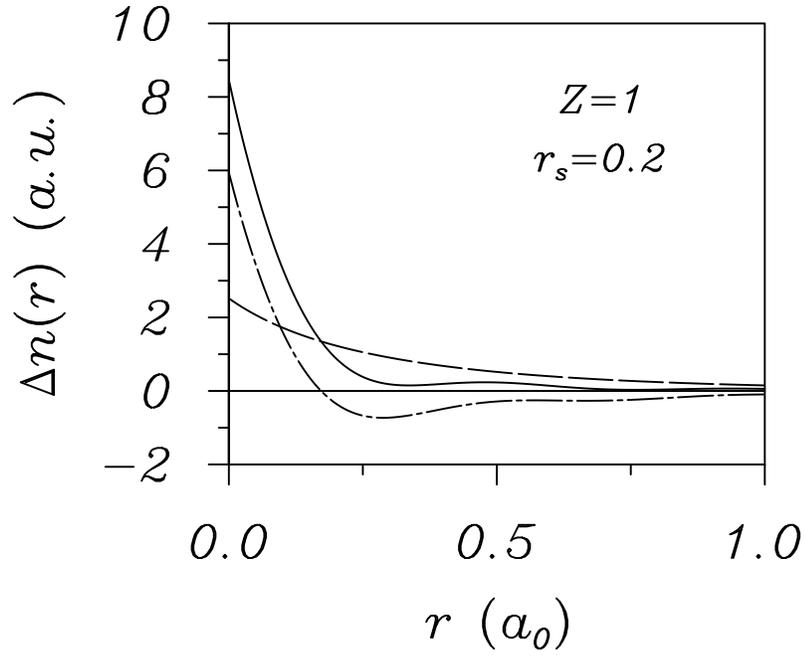}}
 \caption{Screening charge density as a function of position for $Z=1$
and $r_s = 0.2$. The dashed curve is the bound state contribution, the
chain curve is the continuum state contribution and the solid curve is
the total.
 }   
\label{fig14}
\end{figure}

\begin{figure}
\centering
\scalebox{0.6}
{\includegraphics[0.5in,1.5in][8.5in,8.0in]{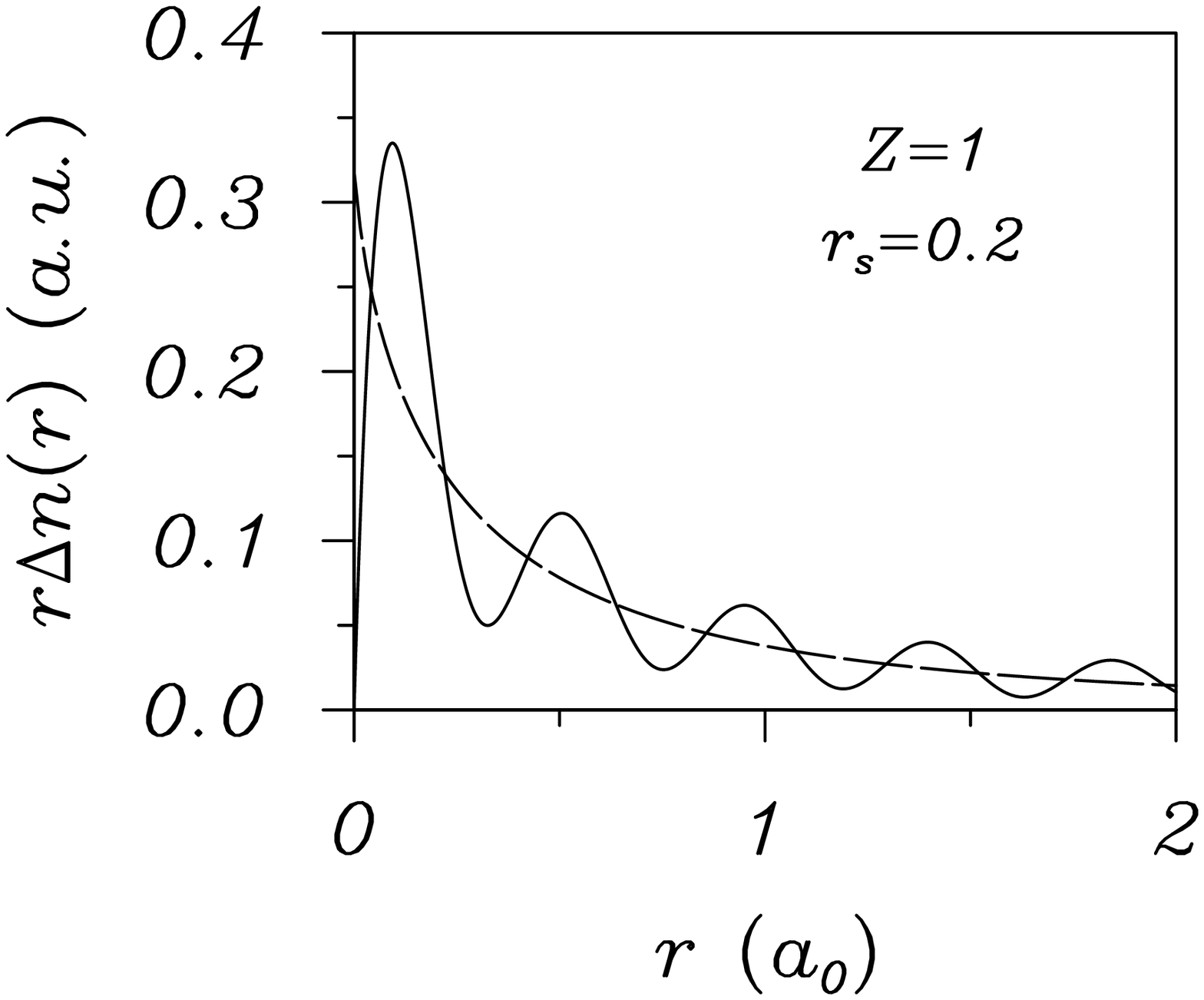}}
 \caption{As in Fig.~14, but plotting the screening charge density
 as $r\Delta n(r)$ {\it vs.} $r$. The dashed curve is the limiting ($r_s
\to 0$) Thomas-Fermi density.
 }   
\label{fig15}
\end{figure}


\begin{references}
                                                            
\bibitem{ando82} T. Ando, A. B. Fowler and F. Stern, Rev. Mod. Phys.
{\bf 54}, 437 (1982).
\bibitem{echenique04} P. M. Echenique, R. Berndt, E.V. Chulkov, Th. Fauster,
A. Goldman and U. H\"ofer, Surf. Sci. Rep. {\bf 52}, 219 (2004).
\bibitem{andrei97} {\it Two-Dimensional Electrons on Cryogenic
Substrates}, ed. by E. Andrei (Kluwer Academic Press, Dordrecht, 1997).
\bibitem{layered} M. S. Dresselhaus and G. Dresselhaus, Adv. Phys.
{\bf 51}, 1 (2002); J. Singleton and C. Mielke, Contemp. Phys. {\bf
43}, 63 (2002).
\bibitem{fletcher90} R. Fletcher, E. Zaremba, M. D'Iorio, C.T. Foxon and
J.J. Harris, Phys. Rev. B {\bf 41}, 10649 (1990).
\bibitem{nagao01} T. Nagao, T. Kildebrandt, M. Henzler and S. Hasegawa,
Phys. Rev. Lett. {\bf 86}, 5747 (2001).
\bibitem{murphy95} S. Q. Murphy, J.P. Eisenstein, L.N. Pfeiffer and K.W. West,
Phys. Rev. B {\bf 52}, 14825 (1995).
\bibitem{stern67} F. Stern and W. E. Howard, Phys. Rev. {\bf 163},
816 (1967).
\bibitem{vinter82} B. Vinter, Phys. Rev. B {\bf 26}, 6808 (1982).
\bibitem{simon76} B. Simon, Ann. Phys. (N.Y.) {\bf 97}, 279 (1976).
\bibitem{zaremba77} E. Zaremba, L.M. Sander, H.B. Shore, and J.H. Rose,
J. Phys. F {\bf 7}, 1763 (1977).
\bibitem{zaremba03} E. Zaremba, I. Nagy and P.M. Echenique, Phys. Rev. Lett.
{\bf 90}, 046801 (2003).
\bibitem{fetter73} A. Fetter, Ann. Phys. (N.Y.) {\bf 81}, 367 (1973).
\bibitem{horing87} N.J.M. Horing, H.C. Tso, and G. Gumbs, Phys. Rev. 
B {\bf 36}, 1588 (1987).
\bibitem{bergara97} A. Bergara, I. Nagy, and P.M. Echenique, Phys. Rev.
B {\bf 55}, 12864 (1997).
\bibitem{wang95} Y.-N. Wang and T.-C. Ma, Phys. Rev. B {\bf 52}, 
16395 (1995).
\bibitem{bret93} A. Bret and C. Deutsch,  Phys. Rev. E {\bf 48}, 2994
(1993).
\bibitem{hu88} C.D. Hu and E. Zaremba, Phys. Rev. B {\bf 37}, 9268
(1988).
\bibitem{bergara99} A. Bergara, J.M. Pitarke, and P.M. Echenique, Phys. 
Rev. B {\bf 59}, 10145 (1999).
\bibitem{nagy95} I. Nagy, Phys. Rev. B {\bf 51}, 77 (1995).
\bibitem{wang97} Y.-N. Wang and T.-C. Ma, Phys. Rev. B {\bf 55}, 
2087 (1997).
\bibitem{sjolander72} A. Sj\"olander and M.J. Stott, Phys.
Rev. B {\bf 5}, 2109 (1972).
\bibitem{krakovsky95} A. Krakovsky and J. K. Percus, Phys. Rev. B {\bf
52}, R2305 (1995).
\bibitem{nagy89} I. Nagy, A. Arnau, P.M. Echenique, and E. Zaremba, 
Phys. Rev. B {\bf 40}, 11983 (1989).
\bibitem{tanatar89} B. Tanatar and D. M. Ceperley, Phys. Rev. B 
{\bf 39}, 5005 (1989).
\bibitem{adhikari86} S. K. Adhikari, Am. J. Phys. {\bf 54}, 362
(1986).
\bibitem{footnote2} The screened potential is of course singular at the 
origin for an impurity point charge. To make the
argument precise one can imagine the impurity having a small but finite 
radius so that the potential is bounded. This 
will have little effect on the screening density in most regions of space.
In 2D the same effect can be achieved by displacing the point charge a
small distance from the plane of the 2DEG.
\bibitem{shore77} H.B. Shore, J.H. Rose, and E. Zaremba, Phys. Rev. B
{\bf 15}, 2858 (1977).
\bibitem{richter89} J. Richter, Phys. Rev. B {\bf 39}, 6268 (1989); E. 
Zaremba, Phys. Rev. B {\bf 44}, 1379 (1991).
\bibitem{ghazali95} A. Ghazali and A. Gold, Phys. Rev. B {\bf 52},
16634 (1995).
\bibitem{footnote1} We remark that a negative sign sometimes appears
in Eq.~(\ref{epsRPA}) 
corresponding to a different choice for the sign of
the response function $\chi_0(q)$.
\bibitem{nagy99} I. Nagy, Phys. Rev. B {\bf 60}, 4404 (1999).
\bibitem{bonig89} L. B\"onig and K. Sch\"onhammer, Phys. Rev. B {\bf
39}, 7413 (1989).
\bibitem{echenique91} P.M. Echenique, A. Arnau, M. Pe\~nalba, and I.
Nagy, Nucl. Instrum. Methods Phys. Res. B {\bf 56/57}, 345 (1991).
\bibitem{krakovsky95b} A. Krakovsky and J. K. Percus, Phys. Rev. B {\bf
52}, 7901 (1995).
\bibitem{butler62} D. Butler, Proc. Phys. Soc. {\bf 80}, 741 (1962).
\bibitem{kohn65} W. Kohn and C. Majumdar, Phys. Rev. {\bf 138}, A1617
(1965).
\bibitem{vanzyl99}
B. P. van Zyl and E. Zaremba, Phys. Rev. B {\bf 59}, 2079 (1999).

\end{references}
\end{document}